\documentclass[usegraphicx,useAMS,usenatbib]{mn2e}

\usepackage{amssymb}
\usepackage{aas_macros}
\usepackage{array}
\usepackage{xspace}
\usepackage{amsmath}
\usepackage{epsfig}
\usepackage{graphicx}
\usepackage{hyperref}
\usepackage{aas_macros}
\usepackage{subfigure}
\usepackage[T1]{fontenc}
\usepackage{color}
\usepackage{float}
\usepackage{bm}

\def\nW{$\mathrm{nWm^{-2}sr^{-1}}$}
\def\mic{$\mu {\rm m}$}
\def\gsim{\gtrsim}

\begin{document}

\title{ AKARI Near-Infrared Background Fluctuations Arise from Normal Galaxy Populations }

\author[K. Helgason, E. Komatsu]
{K. Helgason\thanks{E-mail: kari@mpa-garching.mpg.de }$^1$, E. Komatsu$^{1,2}$ \\
$^1$Max Planck Institute for Astrophysics, Karl-Schwarzschild-Str. 1, 85748 Garching \\
$^2$Kavli Institute for the Physics and Mathematics of the Universe (Kavli IPMU, WPI), \\
Todai Institutes for Advanced Study, the University of Tokyo, Kashiwa 277-8583 Japan }

\maketitle

\begin{abstract}

We show that measurements of the fluctuations in the near-infrared background (NIRB) from the {\it AKARI} satellite  can be explained by faint galaxy populations at low redshifts. We demonstrate this using reconstructed images from deep galaxy catalogs (HUGS/S-CANDELS) and two independent galaxy population models. In all cases, we find that the NIRB fluctuations measured by {\it AKARI} are consistent with faint galaxies and there is no need for a contribution from unknown populations. We find no evidence for a steep Rayleigh-Jeans spectrum for the underlying sources as previously reported. The apparent Rayleigh-Jeans spectrum at large angular scales is likely a consequence of galaxies being removed systematically to deeper levels in the longer wavelength channels.

\end{abstract}

\begin{keywords}
 cosmology: diffuse radiation, large-scale structure of universe -- infrared: diffuse background, galaxies
\end{keywords}

\section{ Introduction }

A significant portion of the diffuse Near-Infrared Background (NIRB) has been resolved into point sources, although the actual amount is still uncertain. This uncertainty is mostly due to the disagreement between determinations of the absolute flux level of the NIRB. Whereas direct measurements of the NIRB report a large flux above that of the integrated light of resolved galaxies \citep[e.g.,][]{DwekArendt98, Matsumoto05}, the levels inferred by $\gamma$-ray absorption studies are much lower \citep{Ackermann12,HESS13,BiteauWilliams15,MAGIC16}.

Additional information can be obtained from spatial fluctuations in the unresolved NIRB. Several teams have measured fluctuations after carefully removing resolved sources from the maps. Measurements in the $1-2$\mic\ range have been made using {\it 2MASS} \citep{Kashlinsky02,Odenwald03}, {\it HST}/NICMOS \citep{Thompson07a,Thompson07b}, {\it CIBER} \citep{Zemcov14}, and {\it HST}/WFC3 \citep{MitchellWynne15}, and in the $2-5$\mic\ range using {\it Spitzer}/IRAC \citep{KAMM1,KAMM2,Cooray12b,Kashlinsky12} and {\it AKARI}/IRC \citep{Matsumoto11,Seo15}. Measurements from the infrared camera (IRC) onboard the {\it AKARI} satellite are particularly important as its wavelength coverage bridges the gap between {\it Spitzer}/IRAC ($>3$\mic) and that of {\it HST} and {\it CIBER} ($< 2$\mic). The {\it AKARI} data therefore carries much weight in constraining the nature of the sources producing the fluctuation signal.

At small angular scales, the power spectrum of the fluctuation signal is mostly consistent with shot noise expected from undetected galaxy populations. The signal tends to rise above the shot noise towards intermediate and large scales, with an amplitude that is thought to be inconsistent with faint galaxies \citep{Helgason12} (H12 hereafter) and other foregrounds such as Diffuse Galactic Light (DGL) or Zodiacal Light (ZL) \citep{Pyo12, Tsumura13ZL,Tsumura13DGL, Arai15, Arendt16}. This has often been referred to as the ``excess'' NIRB fluctuations, the origin of which is currently debated \citep{KAMM3,Fernandez10,Cooray12b,Yue13b,Zemcov14,Gong15,Helgason16,Yue16b}.

Using {\it AKARI}, \citet{Matsumoto11} (M11 hereafter) measured fluctuations in a circular $10^\prime$ region at the north ecliptic pole (NEP) reaching a limiting magnitude of $\sim 23$. \citet{Seo15} (S15 hereafter) carried out a similar measurement in a larger region around the NEP reaching angular scales of $35^\prime$ and removing sources down to $\sim 22$ magnitude. Both studies detected fluctuations at large scales and claimed they could not be from faint galaxies or other foregrounds. In addition, M11 argued that the spectral energy distribution (SED) of this excess signal was a steep Rayleigh-Jeans type spectrum, consistent with hot PopIII-like objects.

\begin{figure}
\centering
      \includegraphics[width=0.49\textwidth]{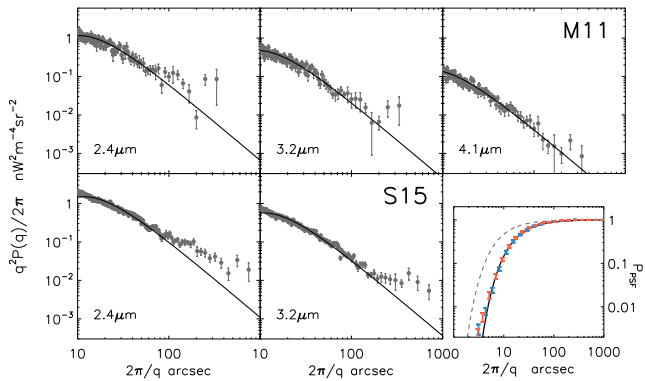}
      \caption{ Power spectra measured by the {\it AKARI}/IRC. The upper panels show the measurements of M11 whereas the lower panels the measurements of S15. The latter shows the raw, uncorrected power spectra to illustrate the PSF suppression. The black line is a two-component fit to the data in the $10^{\prime\prime}-100^{\prime\prime}$ range, $P= P_{\rm SN} + P_{\rm IN}$, where $P_{\rm SN}$ is a shot noise-like component suppressed by the PSF on small scales and $P_{\rm N} = const.$ is an instrument noise-like component. The panel in the lower right corner shows the normalized power spectrum of the PSF. The data points are from the PSF determination in \citet{Seo15} at 2.4\mic\ (blue) and 3.2\mic\ (red) which we have fitted with an exponential profile (black line). The dashed line shows the {\it Spitzer}/IRAC PSF (ch1). }
      \label{fig_beam}
\end{figure}

Figure \ref{fig_beam} shows the power spectra measured by {\it AKARI} \citep{Matsumoto11,Seo15}\footnote{We show the raw fluctuation data of \citet{Seo15} which are not corrected for the Point Spread Function (PSF), mask, or map making transfer function.} and a two-component noise model fitted to them. A shot noise-like component, $P_{\rm SN}$, is suppressed by the PSF on small scales, while an instrument noise-like component, $P_{\rm IN}$, is unaffected by the PSF. When compared with M11 (upper panels), there is little evidence for any fluctuations in excess of the noise power; however, an excess is seen in S15 at large angular scales (lower panels).

In this paper, we use three independent methods to carefully quantify the contribution of faint galaxies to the {\it AKARI} measurement. We use the Planck first-year cosmological parameters to match the semi-analytic models described in Section \ref{sec2} \citep{Planck1}. All magnitudes are in the AB system \citep{OkeGunn83}.

\section{ Fluctuations from galaxy populations } \label{sec2}
\subsection{ Reconstructed catalog images }

To compute the power spectrum of the known galaxy populations undetected by {\it AKARI}, we use catalogs from deep survey images of the UDS and GOODS-S fields in the three {\it AKARI}/IRC channels. We combine two public catalogs: Spitzer-Cosmic Assembly Deep Near-Infrared Extragalactic Legacy Survey (S-CANDELS) and the Hawk-I UDS and GOODS Survey (HUGS). S-CANDELS is a {\it Spitzer}/IRAC program \citep{Ashby15} providing deep coverage in the five CANDELS fields \citep{Grogin11,Koekemoer11}. HUGS is a deep imaging survey in two of the CANDELS fields, UDS and GOODS-S, executed with the Hawk-I imager at the {\it VLT} in the $Y$ and $K$ bands \citep{Fontana14}. We use the final Hawk-I release\footnote{http://www.astrodeep.eu/hugs-data-release/} which is cataloged together with all the multi-band CANDELS data in GOODS-S. The corresponding catalog for the UDS field is described in \citet{Galametz13}\footnote{http://candels.ucolick.org/data\_access/UDS.html}. S-CANDELS and HUGS reach a depth of $\sim$26-27 magnitude which is well beyond the depth of {\it AKARI} NEP deep field, detecting more than $10$ times more galaxies per unit area.

We first create empty images with a size determined by the S-CANDELS coverage of the two fields and a pixel size of $0.6^{\prime\prime}$. The catalogued sources are inserted in positions given by the catalog entries ({\tt RA},{\tt dec}) as single-pixel delta functions with a flux density $S_\nu$ in $\mu$Jy given by {\tt HAWKI\_Ks\_FLUX}, {\tt IRAC\_CH1\_FLUX} and {\tt IRAC\_CH2\_FLUX}. The flux density is then converted to a surface brightness, $I_\nu$ (Jy/sr) for the pixel solid angle, which is further transformed into $\nu I_\nu$ (${\rm nW~m^{-2}sr^{-1}}$). For all galaxies present in all three channels, we linearly interpolate their flux to the {\it AKARI}/IRC center wavelengths of 2.4\mic, 3.2\mic\ and 4.1\mic.

To avoid overdensities of sources due to non-uniform coverage (e.g., in HDF12 is a deep subregion of GOODS-S), we remove all sources that are fainter than the {\it brightest limiting magnitude} at any location of the field according to the catalog entries {\tt
Limiting\_Magnitude\_Khawki} and {\tt Limiting\_Magnitude\_irac1(2)}. This value is around 27 AB in our final region but we conservatively choose $m_{\rm max} =26.5$ AB in all images to ensure uniformity. This choice does not affect our results unless $m_{\rm max} < 25$. Our final images contain all $<26.5$ magnitude sources in the regions that are $21.6^\prime \times 7.9^\prime$ (UDS) and $8.3^\prime \times 10.1^\prime$ (GOODS-S) centered at ${\rm (RA,dec)}=(34.41,-5.20)$ and $(53.11,-27.77)$ respectively.

\begin{figure*}
\centering
      \includegraphics[width=0.98\textwidth]{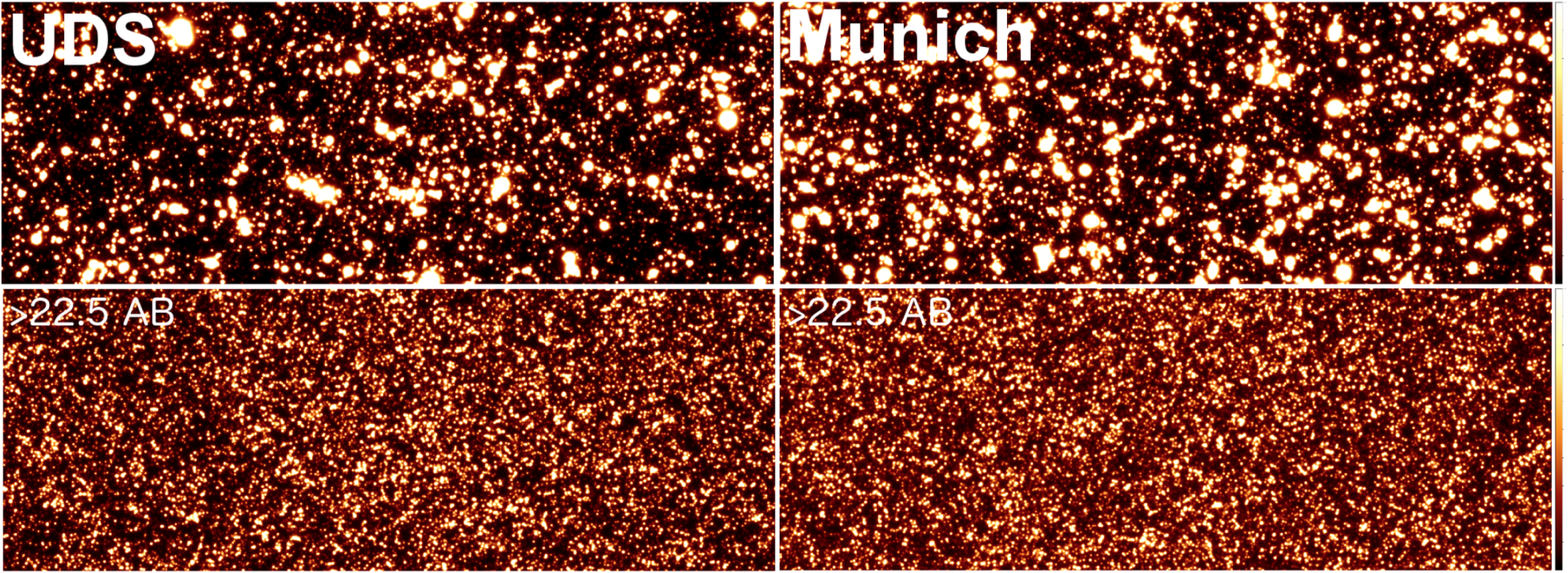}
      \caption{ {\it Left}: A $21.6^\prime\times 7.9^\prime$ region reconstructed from the CANDELS-UDS catalog and convolved with the {\it AKARI}/IRC PSF. The top image contains all sources and the color scale is linear in the [0,20]\nW\ range. The bottom image we have removed sources brighter than 22.5 magnitude where the color scale is in the [0,4]\nW\ range. {\it Right}: An image reconstructed from the Millennium light-cones using the semi-analytical Munich model, cropped to the same region and displayed using the same color scale and ranges.  }
      \label{fig_im}
\end{figure*}

We convolve the final images with a circularly-averaged PSF of the {\it AKARI}/IRC. S15 determined the power spectrum of the PSF by stacking point sources (see the bottom right panel in Figure \ref{fig_beam}). We fit an exponential profile of the form $P_{\rm PSF}(q)= {\rm exp}\left[-(q/q_0)^\gamma\right]$ to the PSF power spectrum data, resulting in the best-fit values of $\gamma=1.2$ and ${\rm log}~q_0=4.87$. As the differences of the PSF in channels 1 and 2 are insignificant, we adopt the same PSF in all channels. The PSF-convolved images are shown in the left panels of Figure~\ref{fig_im}.

\begin{figure*} 
\centering
\begin{minipage}[t]{.48\textwidth}
\includegraphics[width=0.99\textwidth]{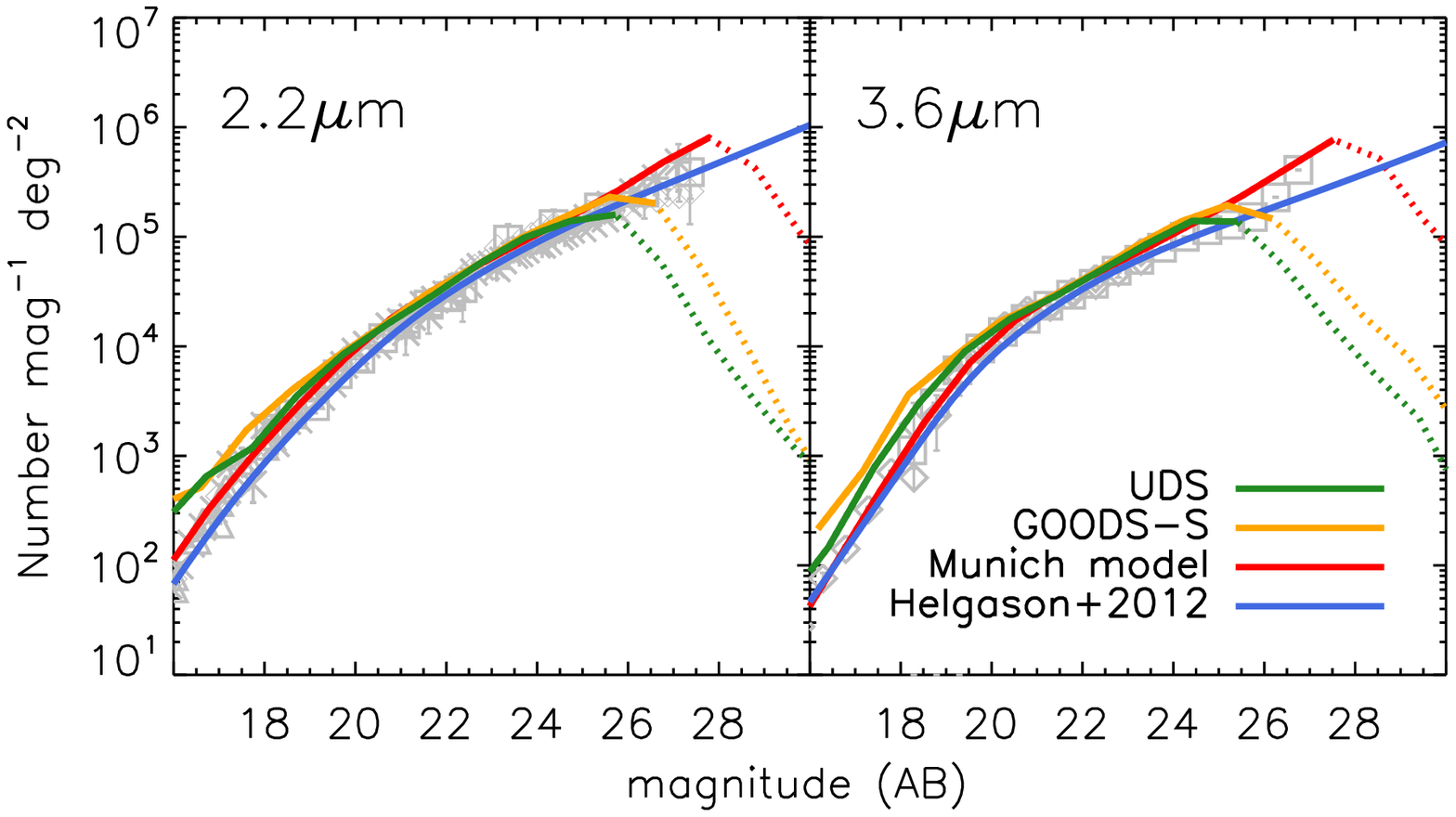}
      \caption{ Galaxy number counts from our models (solid lines) at 2.2\mic\ (left) and 3.6\mic\ (right). The dotted lines indicate counts where they are incomplete due to detection limit or simulation resolution. The observational data are shown in grey (in addition to \citet{Fontana14} and \citet{Ashby15}, see \citet{Helgason12} and references therein).}
\label{fig_counts}
\end{minipage}
  \hfill
\begin{minipage}[t]{.48\textwidth}
\centering
\includegraphics[width=0.99\textwidth]{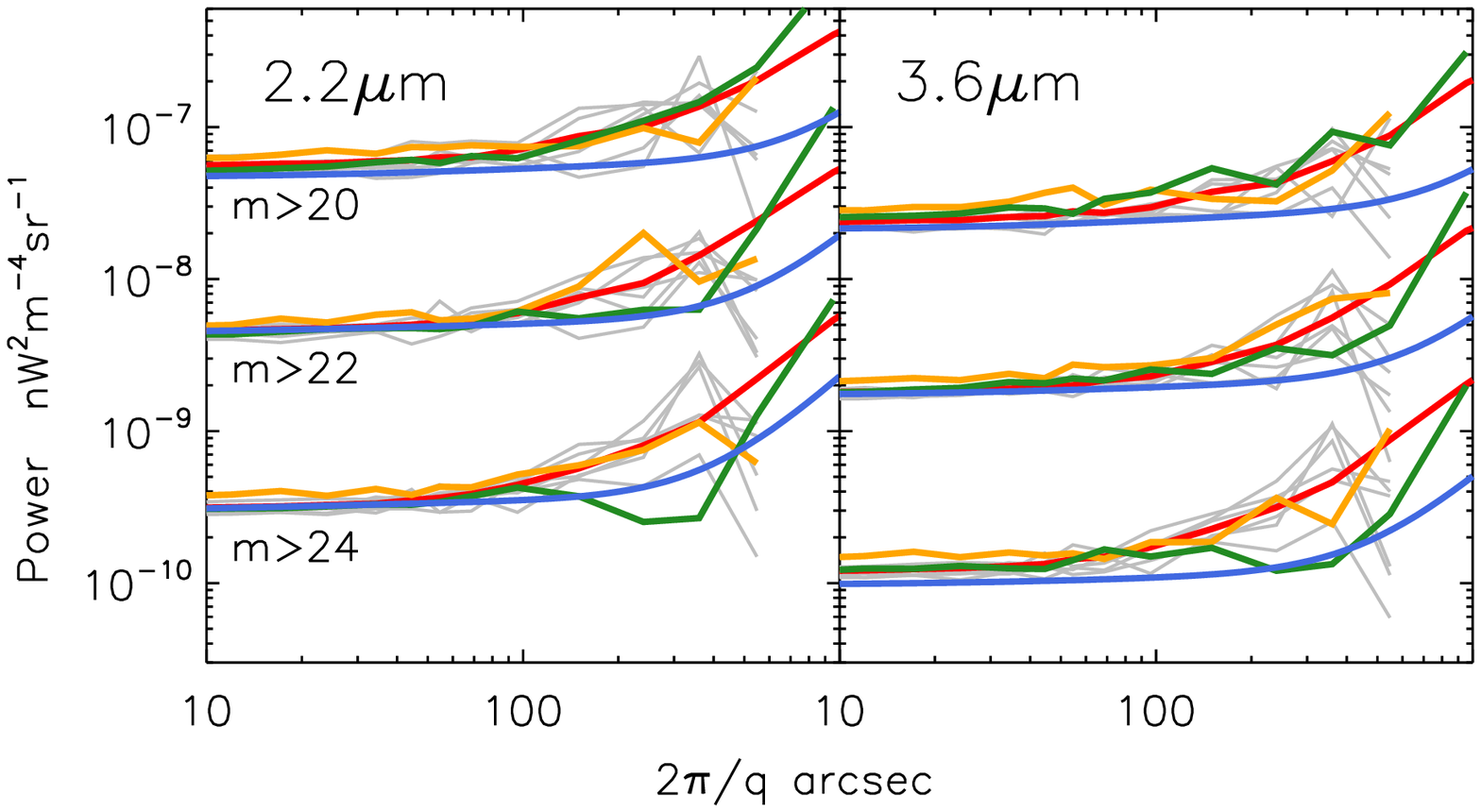}
      \caption{Power spectra from our models. The color scheme is the same as in Figure~\ref{fig_counts}. The three families of lines correspond to sources removed down to magnitude 20, 22 and 24 (top to bottom). The grey curves show six different realizations constructed from the Millennium light cones, cropped to GOODS-S size to illustrate a field-to-field variance. The red line shows the average of these for the full-sized fields $35^\prime\times 24^\prime$. The clustering from the H12 model is systematically lower (see text).  }
\label{fig_power}
\end{minipage}
\end{figure*}

\subsection{ The Munich model }

Next, we construct simulated images of galaxy populations using the Munich galaxy formation model, a full semi-analytic galaxy evolution model coupled with the Millennium simulation \citep{Springel05,Henriques15}. The latest version of the Munich model matches the existing observations of galaxy luminosity functions and the two-point correlation functions of galaxies fairly accurately (see \citet{Henriques15} and \citet{vanDaalen16} for details).

We make use of the publicly available light-cones in the Millennium database with galaxy magnitudes pre-computed for filters in various observatories. The light-cone construction is described in \citet{Henriques12} and assumes Planck 1-year cosmology. The projection is a circular field with a $2^\circ$ diameter out of which we use a rectangular $35^\prime \times 10^\prime$ region in the center.

To capture a field-to-field variance, we use six independent light-cones from the public Millennium database {\tt Henriques2015a.cones.MRscPlanck1\_BC03\_0ij} (with {\tt ij}=01--06) which employ stellar population synthesis models of \citet{BruzualCharlot03}\footnote{We also explored a stellar population synthesis model of \citet{Maraston05}, finding little change to our results.}. We use extinction-corrected magnitudes provided at 2.2\mic\ ({\tt Ks}), 3.6\mic\ ({\tt i1}) and 4.5\mic\ ({\tt i2}) and interpolate between them to get galaxy magnitudes at the centers of the {\it {\it AKARI}} channels.

The catalogued galaxies are inserted in image positions given by ({\tt ra},{\tt dec}) as single-pixel delta functions with their magnitude converted into $\nu I_\nu$ (${\rm nW~m^{-2}sr^{-1}}$) and convolved with the PSF of {\it AKARI}/IRC. We do not account for the shape or sizes of extended sources as they are generally much smaller than the FWHM of the {\it AKARI} PSF. The PSF-convolved images are shown in the right panels of Figure~\ref{fig_im}.

\subsection{ Helgason et al. 2012 }

Finally, we use the galaxy population model of \citet{Helgason12} (H12, hereafter). This empirical model reconstructs the light cone using a library of measured galaxy luminosity functions in UV/optical/NIR that together cover $0<z<8$. The model provides the NIRB flux production as a function of redshift for a desired source removal threshold, $m_{\rm lim}$. NIRB fluctuations are then calculated based on a halo model with a commonly adopted halo occupation distribution (HOD). We refer to \citet{Helgason12} for details.

 \section{ Results } \label{sec3}

Figure \ref{fig_counts} shows the galaxy counts of all our final data products as well as those of the observational data at 2.2\mic\ and 3.6\mic. We find good agreement in the range of interest ($20<m<25$) before incompleteness sets in at $\gsim 26$. Slightly elevated counts at brighter magnitudes in USD and GOODS-S are likely due to star contamination but this does not affect our results. The counts from the Munich model are ``complete'' out to $\sim$28 mag at which they become limited by the finite resolution of the Millennium simulation.

Figure \ref{fig_power} shows the angular power spectra of galaxies after sources brighter than $m=20$, $22$, and $24$ have been removed. This is calculated directly from the images using $P(q)=\langle |\Delta_q|^2 \rangle$ where $\Delta_q$ is the two dimensional Fourier transform of the image calculated using the FFT and $q$ is the angular wavenumber. The brackets $\langle ~ \rangle$ denote the average over the azimuthal angle in Fourier space. We find that all the power spectra are in reasonable agreement with the exception of the H12 model which falls systematically below the others. This is due to inaccuracy of the H12 model; 
in particular the choice of the minimum mass of a halo that hosts a galaxy (H12 chose $M_{\rm min}=10^9M_\odot$,
which was too low).

\begin{figure}
\centering
      \includegraphics[width=0.37\textwidth]{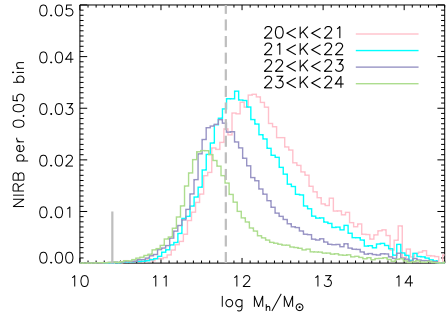}
      \caption{ The NIRB flux production as a function of the central host halo mass in the semi-analytical Munich model (in \nW). The distribution is shown in four different magnitude bins down to 24 AB. At these magnitudes, the unresolved NIRB is dominated by galaxies just beyond the detection threshold. This shows how the dominant host halo mass, hence clustering of the NIRB fluctuation, is sensitive to the magnitude limit of the survey. The vertical dashed line shows the $M_{\rm min}$ we have chosen for the new H12 model that reproduces the {\it AKARI} measurements with $m_{\rm lim}=$ 21 to 24 AB. The vertical solid line shows the approximate resolution limit of the Millennium simulation, well below the relevant halo masses dominating the NIRB at these levels.    }
      \label{fig_hod}
\end{figure}

Figure \ref{fig_hod} shows the relative NIRB contribution from faint galaxies as a function of the host halo mass in the Millennium light-cones. This suggests that most of the NIRB comes from halos $>10^{11}M_\odot$ and that the average occupation, hence clustering in the fluctuations, is quite sensitive to the magnitude limit of surveys. We can bring the H12 model into agreement with the other models and the data by setting ${\rm log} M_{\rm min}=11.8$, which increases the galaxy bias, elevating the large scale fluctuations by a factor of $\sim 2$.

\begin{figure*}
\centering
      \includegraphics[width=0.98\textwidth]{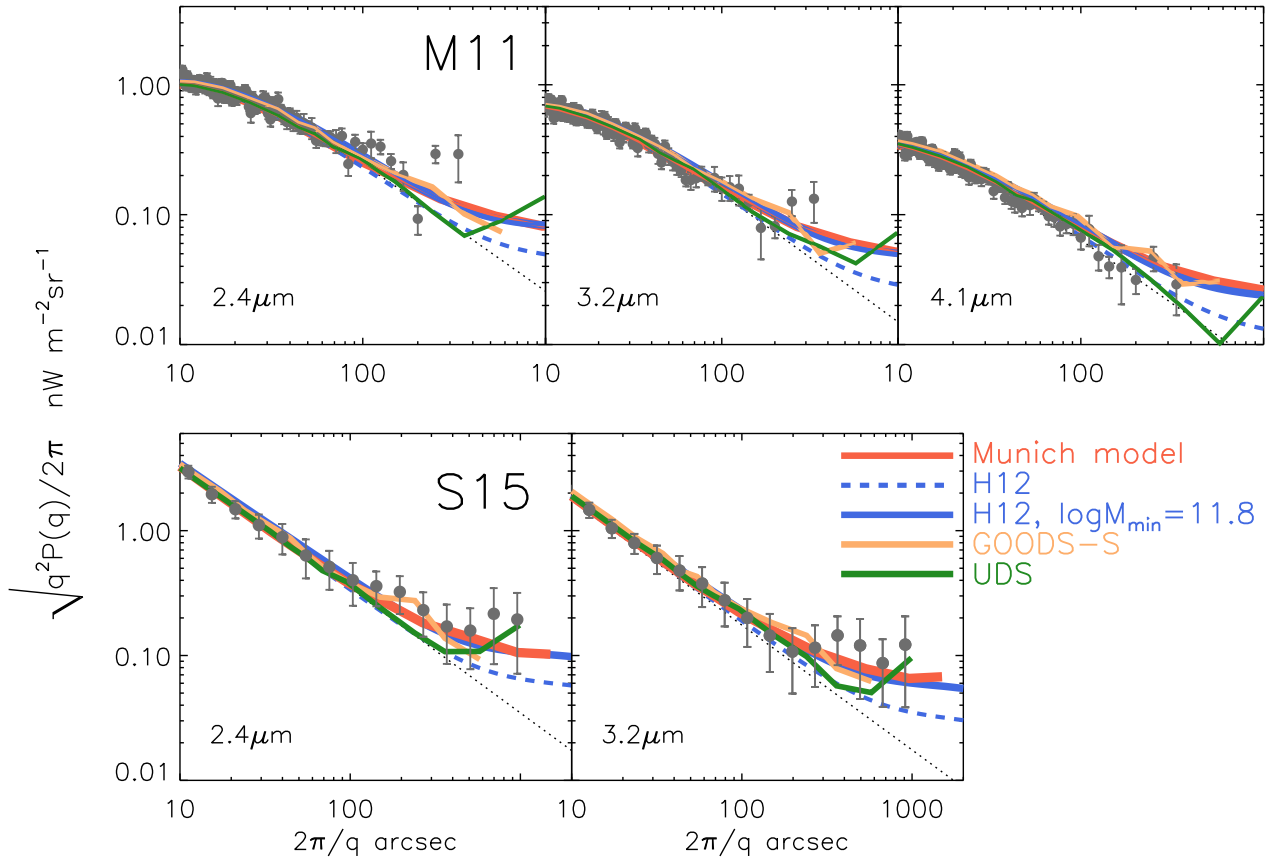}
      \caption{ NIRB fluctuations of all our model images compared with the {\it AKARI} measurements of \citet{Matsumoto11} (upper panels) and \citet{Seo15} (lower panels): the Munich model (red), the original \citet{Helgason12} model (blue), the \citet{Helgason12} model with $M_{\rm min}$ modified (dashed blue), and the reconstructed GOODS-S and UDS images (orange and green respectively). The dotted lines show a pure shot noise power for comparison.  }
      \label{fig_results}
\end{figure*}

Figure \ref{fig_results} presents the main results of this paper: the power spectra of both our empirically constructed images and models are in a reasonable agreement with the {\it AKARI} fluctuation data. For all the lines we choose the same limiting magnitude: 22.5, 22.8 and 23.5 at 2.4, 3.2 and 4.1\mic\ respectively for M11 and 22.0 and 22.4 at 2.4 and 3.2\mic\ respectively for S15. The shot noise dominating the small scale fluctuations is in good agreement among the different models despite all having the same $m_{\rm lim}$. On larger scales, $>100^{\prime\prime}$, all the power spectra exhibit clustering above the noise which is sufficient to explain the data.

The magnitude limits giving the best match to the M11 data are roughly 0.4 mag brighter than the ones quoted in M11. We also find roughly 0.6 mag fainter magnitude limits for S15 than the ones quoted in their paper. This is not unusual as these limits are estimated based on the amplitude of the shot noise which in turn relies on assumptions on the surface density of unresolved galaxies.

S15 estimated the contribution of unresolved galaxies using ground-based observations of the NEP deep field \citep{Oi14} and found fluctuations that are 2.8 times lower than the measured signal. This is notably lower than what we are finding. The difference could be due to the NEP observations being only 50\% complete  at $K_s=22.3$ AB such that a large portion of the faint galaxies would be missing, resulting in a lower fluctuation power (our catalogs are complete out to 25.5 AB). This is supported by the fact that S15 reproduce their shot noise with a limiting magnitude of 21.5 AB whereas we find the same shot noise levels at $m_{\rm lim}=22$ AB, indicating a higher surface density of unresolved galaxies.

Although we find that {\it AKARI} fluctuations can be explained by normal galaxies, the same is not true for {\it Spitzer}/IRAC measurements at similar wavelengths which still show fluctuations in excess of what can be attributed to faint galaxies. Because of the difference in the survey depth however, this does not mean that the {\it AKARI} measurements are in conflict with {\it Spitzer}. A fluctuation excess of the amplitude seen by {\it Spitzer}/IRAC at 3.6\mic, scaled to 3.2\mic\ using a $\lambda I_\lambda \sim \lambda^{-3}$ SED, can be accommodated within the uncertainties on top of the contribution from normal galaxies. But no such component is required based on the {\it AKARI} data alone.

M11 argued that the fluctuation SED was steep, close to Rayleigh-Jeans, $\lambda I_\lambda \propto \lambda^{-3}$, characteristic of hot PopIII stars. While it is true that the amplitude of the large scale fluctuations compared across the three channels shows a Rayleigh-Jeans type SED, the small-scale fluctuations, which are almost certainly dominated by normal galaxies, also exhibit such a steep SED. But with a uniform source removal in all channels, the spectrum of the NIRB from galaxies should be much shallower, or roughly $\lambda I_\lambda \propto \lambda^{-1.0}$. The only way this can happen is if sources are removed to different levels in different wavelength channels. Indeed, the limiting magnitudes increase by $\Delta m \simeq$ 1.0 AB from 2.4 to 4.1\mic\ making the SED appear steeper. This may be due to the source cleaning (PSF subtraction) algorithm used by M11 and S15, which can proceed to deeper levels in the more sensitive channels.

Measurements from {\it CIBER} found an apparent continuation of this Rayleigh-Jeans SED to shorter wavelengths (1.1 and 1.6\mic) but other measurements at the same wavelengths, but at different depths, from {\it 2MASS} and {\it HST}/WFC3 do not support this showing widely different amplitudes.

\section{ Conclusions } \label{sec4}

We have used reconstructed survey images and galaxy population models to examine the contribution of faint galaxies to the NIRB fluctuations measured with {\it AKARI}. We find that the data are consistent with faint galaxies and there is no need for a contribution from unknown populations based on the {\it AKARI} data alone. Additionally, we find no evidence for a Rayleigh-Jeans type SED for the underlying sources. We are able to fit the fluctuations at all wavelengths and angular scales with normal populations. The apparent Rayleigh-Jeans slope is likely a consequence of galaxies removed systematically to deeper levels in the longer wavelength channels. These results do not rule out a fluctuation excess of the amplitude seen in {\it Spitzer}/IRAC measurements at similar wavelengths. At the depths of the {\it AKARI}, this component would be sub-dominant compared to the normal galaxies.

\section*{ACKNOWLEDGMENTS}

KH was supported by the European Unions Seventh Framework Programme (FP7-PEOPLE-2013-IFF) under grant agreement number 628319-CIBorigins. EK was supported in part by JSPS KAKENHI Grant Number JP15H05896.


\end{document}